\documentstyle[prl,aps,amssymb,amsmath,twocolumn]{revtex}
\begin{document}
\draft
\title{No-nonsense Casimir force}
\author{Andrzej Herdegen\cite{AH}}
\address{Institute of Physics, Jagiellonian University, ul.\ Reymonta 4,
30-059 Krak\'ow, Poland}
\maketitle
\begin{abstract}
Two thin conducting, electrically neutral, parallel plates forming an
isolated system in vacuum exert attracting force on each other, whose
origin is the quantum electrodynamical interaction. This theoretical
hypothesis, known as Casimir effect, has been also confirmed
experimentally. Despite long history of the subject, no completely
convincing theoretical analysis of this effect appears in the literature.
Here we discuss the effect (for the scalar field) anew, on a revised
physical and mathematical basis. Standard, but advanced methods of
relativistic quantum theory are used. No anomalous features of the
conventional approaches appear. The Casimir quantitative prediction for
the force is shown to constitute the leading asymptotic term, for large
separation of the plates, of the full, model-dependent expression.
\end{abstract}
\vspace*{-.5cm}
\narrowtext
One of the significant visualizations of a quantized field in the early
development of the relativistic quantum theory has been that of an
ensemble of oscillators (see e.g.\ \cite{wen}). One found a close
mathematical relation between a quantum field and an infinite collection
of quantum harmonic oscillators, each characterized by its characteristic
frequency $\omega_\alpha$, and taking on one of the energy values
$E_{\alpha,n_\alpha}=\hslash\omega_\alpha(n_\alpha+\frac{1}{2})$. The
collection of the amplitudes of the oscillators reflects the strength of
the field. The lowest energy state of the field, the vacuum state, is
represented by all of the oscillators being in their respective ground
states. The mean value of the field vanishes in this state, but the mean
value of the field strength {\sl squared} does not. This shows, one says,
that vacuum fluctuates, the energy of these fluctuations being equal to
$(\hslash/2)\sum_\alpha
\omega_\alpha$ (the ``zero-modes sum'', or ``zero-point energy'', as it
has been christened). This sum is infinite, but, as one is told, the
energy differences is what counts experimentally, not the absolute value.

This early picture has been, of course, superseded by later developments
in the quantum field theory (QFT). In particular, one of the earliest
instances of renormalization, normal ordering of operators, eliminated
the infinite energy of the vacuum state and renormalized it to zero.
Fluctuations of the field strength simply reflect the fact that vacuum is
not an eigenstate of the field (neither is any other state), but they are
not dynamical, and cannot be a possible source of extractable energy.
However, it is an astonishing fact that there does exist an isolated
area, in which the ``zero-modes'' ideology has been in use to date. In a
1948 paper \cite{cas} Casimir considered situation in which two thin
conducting, electrically neutral, parallel plates are placed in the
vacuum. The presence of these plates creates boundary conditions for the
electromagnetic field, which change the characteristic frequencies of
field oscillators to $\omega_\alpha(a)$, depending on the separation of
the plates $a$. The lowest state is now characterized by the energy
$(\hslash/2)\sum_\alpha\omega_\alpha(a)$. This is again infinite, but
Casimir regularized it so as to squeeze a finite result out of it. Minus
derivative of this expression is supposed to give the force between the
plates. His prediction for the force, as is well known, was
$-(\pi^2/240)\hslash c\, a^{-4}$. The paper addressed an interesting and
fundamental quantum phenomenon, and is deservedly regarded as a
pioneering work. The method it used, however, reflects the relatively
early stage of the development of QFT, and today it should not be taken
as a serious base of further research in this field, as it still very
often is (see review articles on the subject \cite{rev}; the Casimir
reasoning is even reported, virtually unchanged, by some respectable
textbooks on modern QFT, see e.g.\ Ref.\cite{itz}). Other methods, based
on the local energy expectation value, have been also used in the present
context \cite{rev}. However, these calculations ignore the algebraic
difficulty which we are going to discuss below. We leave the discussion
of local aspects to a more extensive future publication.

Consider the system investigated by Casimir, a quantum field plus plates,
where for simplicity we take scalar field. Its most satisfactory
description would be achieved, as usually in physics, by constructing a
closed theory of both elements in mutual interaction. Here we leave this
ambitious task aside, we would like to understand first the reasonable
approximate idealization in which the plates are regarded infinitely
heavy (hence, in particular, classical). We want to show that even for
that restricted purpose the usual naive treatment of the concept of
quantum field is not sufficient. Proper care has to be taken with respect
to the algebraic structure of the theory, including the scope of quantum
variables under consideration, their various representations, and the
time evolution of the system. The algebraic aspects of quantum theory,
already stressed by Dirac in his monograph on quantum mechanics, has been
growing in importance, especially with the creation of the Haag--Kastler
framework for local algebras in quantum physics (see the monograph
\cite{haa}). Today the algebraic approach is the most general and
flexible framework for considering fundamental questions in quantum
physics. We start our analysis by sketching the theory of free scalar
field in the initial conditions formulation, stressing the algebraic
aspects. We use the units with $\hslash=1$, $c=1$.

Let ${\mathcal{L}}_0$ denote the real vector space formed by pairs of
functions on the $3$-space $\left( v(\vec{x}), v_t(\vec{x})\right)$, each
of which is an element of the vector space $\mathcal{D}_{\mathbb{R}}$ of
real, infinitely differentiable functions with compact support
(${\mathcal{L}}_0$ is the direct sum
$\mathcal{D}_{\mathbb{R}}\oplus\mathcal{D}_{\mathbb{R}}$). The elements
of ${\mathcal{L}}_0$ will be denoted by $V\equiv(v, v_t)$, and the vector
arguments suppressed. ${\mathcal{L}}_0$ becomes a symplectic space with
the introduction of the symplectic form
\begin{equation}\label{sym}
 \sigma(V_1, V_2)=\int\left( v_2(\vec{x})v_1{}_t(\vec{x})-
 v_1(\vec{x})v_2{}_t(\vec{x})\right) d^3x\, .
\end{equation}
The real scalar quantized field is a set of elements $\phi(V)$ generating
an algebra by the relations
\begin{equation}\label{alg}
 \phi(V)^*=\phi(V)\, ,\ \left[ \phi(V_1), \phi(V_2)\right] =
 i\sigma(V_1, V_2)\, .
\end{equation}
(More precise formulation expresses the above relations in terms of Weyl
elements $W(V)=\exp i\phi(V)$, in order to avoid domain problems.) The
element $\phi((0,v_t))$ has the interpretation of the field operator
``smeared'' with the test function $v_t(\vec{x})$, and the element
$\phi((v, 0))$ -- the interpretation of the canonical momentum
``smeared'' with the test function $v(\vec{x})$; the elements are
localized in the support of their test functions. The free, massless
evolution of the quantum field is obtained by a simple ``quantization''
of the classical linear evolution determined by the wave equation. Denote
by $h_0$ the square root of the positive operator $h_0^2\equiv -\Delta$
acting on $L^2({\mathbb{R}}^3,d^3x)$. Then the evolution of the initial
conditions for this equation is given by the transformation
\begin{equation}\label{cev}
\begin{split}
 v_0(t)&= \cos(h_0t)\, v + \sin(h_0t)\, h_0^{ -1}\, v_t\, ,\\
 v_0{}_t(t)&= -\sin(h_0t)\, h_0\, v+ \cos(h_0t)\, v_t\,
\end{split}
\end{equation}
(implying $v_0(0)=v$, $v_0{}_t(0)=v_t$). Note that $v_0(t)$ and
$v_0{}_t(t)$ are in $\mathcal{D}_{\mathbb{R}}$, so
 $V_0(t)\equiv(v_0(t),v_0{}_t(t))\in{\mathcal{L}}_0$.
The evolution is a symplectic transformation, that is,
\begin{equation}\label{}
 \sigma(V_1{}_0(t), V_2{}_0(t))=\sigma(V_1, V_2)\, .
\end{equation}
The quantum field evolves according to automorphic map of the algebra
\vspace*{-.2cm}
\begin{equation}\label{qev}
 \alpha_{0t}\,\phi(V)= \phi(V_0(t))\, .
\end{equation}

The next step is the construction of the vacuum representation -- the
unique representation in which the operator of energy has an eigenvector,
and its spectrum is bounded from below. This is achieved, as is well
known, by separating ``positive frequencies'' from ``negative
frequencies'' in the evolution law of the field and interpreting the
coefficients of this two parts as creation and annihilation operators
respectively. More precisely, this amounts to the following. Take the
Hilbert space ${\mathcal{K}}\equiv L^2({\mathbb{R}}^3, d^3x)$ as the
``one-particle space''. By taking multiple direct products of
$\mathcal{K}$, and then forming the direct sum of these products,
construct a standard Fock space $\mathcal{H}$. Denote by $\Omega_0$ the
distinguished normalized vector in $\mathcal{H}$ (``Fock vacuum''), and
by $a(f)$, $a^*(f)$, for $f\in\mathcal{K}$, the usual annihilation and
creation operators in $\mathcal{H}$. The real-linear operator
$j_0:{\mathcal{L}}_0\mapsto \mathcal{K}$ defined by
\begin{equation}\label{j}
 j_0(V)=2^{-1/2}\left( h_0^{1/2}\, v - ih_0^{-1/2}\,
 v_t\right)\,
\end{equation}
extracts the positive frequency part of the evolution:
\begin{equation}\label{pos}
 j_0(V_0(t))= e^{ih_0t}j_0(V)\, ,
\end{equation}
(note that $j_0$ is well defined, as
${\mathcal{D}_{\mathbb{R}}}\subset{\mathcal{D}}(h_0^{1/2})\cap
{\mathcal{D}}(h_0^{-1/2})$). The vacuum representation $\pi_0\left(
\phi(V)\right)\equiv\Phi_0(V)$ is now defined by
\begin{equation}\label{rep}
 \Phi_0(V)= a(j_0(V))+a^*(j_0(V))\, .
\end{equation}
In this representation the evolution is implemented by the unitary
operator $U_0(t)=\exp iH_0t$, with $H_0$ having the interpretation of the
field energy operator, by
\begin{equation}\label{imp}
 U_0(t)\Phi_0(V)U_0(t)^{-1}=\Phi_0(V_0(t))\, .
\end{equation}
If $\{ f_i\}_{i=1}^\infty$ is any (orthonormal) basis of $\mathcal{K}$ in
the domain of $h_0^{1/2}$, then $H_0$ may be represented by
\begin{equation}\label{en}
 H_0=\sum_{i=1}^\infty a^*(h_0^{1/2}f_i)\, a(h_0^{1/2}f_i)\, .
\end{equation}
Hence, in particular, $H_0$ is positive, and $\Omega_0$ is the physical
vacuum: $H_0\Omega_0=0$.

The theory is thus defined, but one should bear in mind three levels of
specialization in the construction: \mbox{Eqs.\ (\ref{sym}--\ref{alg})}
define the algebra, Eqs.\ (\ref{cev}--\ref{qev}) the free evolution, and
Eqs.\ (\ref{j}--\ref{en}) the vacuum representation. One should also
point out that the choice of the basic vector space for the canonical
relations (${\mathcal{L}}_0$ above) is to certain extent flexible, as
long as all consistency conditions are satisfied as above.

Now we can return to our task of investigating Casimir effect. In the
first step one has to define a one-parameter family of time evolutions of
our field algebra, enforced by the presence of the conducting plates at
all possible (but fixed) distances $a$ (we place one of them in the
$x\mbox{-}y$ plane ($z=0$), and another parallel at $z=a$). This should
amount to imitating the steps embodied by Eqs.\ (\ref{cev}--\ref{qev}),
with $h_0$ replaced by the square root $h$ of the positive operator $h^2$
in $L^2({\mathbb{R}}^3,d^3x)$ defined uniquely as $-\Delta$ with
Dirichlet conditions on the plates. Here, however, one encounters a
serious difficulty. The new classical evolution law, Eq.\ (\ref{cev})
with $h$ replacing $h_0$, implies that $v$ has to lie in
${\mathcal{D}}(h)$, the domain of $h$. Now, all functions in
${\mathcal{D}}(h)$ vanish on the plates (${\mathcal{D}}(h)$ is equal to
the direct sum of the three Sobolev spaces $H_0^1$ for each of the three
closed regions into which the whole space is divided by the plates
\cite{tay}). The new evolutions may not be defined on our algebra.
Moreover, any other acceptable choice of the symplectic space will not
solve the problem: with varying separation $a$ one sees that $v$ would
have to vanish in the whole region of interest, making the theory
trivial. Physically this means that the idealization of sharp Dirichlet
conditions at variable positions is unphysical, at least in the
approximation of heavy, classical plates. No traditional approach is able
to clarify the source of this difficulty. Trying to ignore the
difficulty, one is bound to encounter infinities of physical, and not
only technical, origin.

The only possible solution is choosing some other model for the
interaction with the plates, some ``softened'' version of the Dirichlet
condition; this ``softening'' will affect the dynamics in the
$z$-direction. Moreover, one should also expect difficulties coming from
the infinite extension of the plates (they would not appear, e.g., for
spherical shells). This is, however, not serious, as we are interested in
quantities (e.g.\ force) per unit area of the plates, so we can
approximate by large, but finite extension plates (taking limit at an
appropriate point). What we propose, therefore, is the following. Put
$h_\perp^2=-\partial_x^2-\partial_y^2$ on $L^2((-L_x/2, +L_x/2)\times
(-L_y/2, +L_y/2), dx\, dy)$ with Dirichlet conditions at $x=\pm L_x/2$,
$y=\pm L_y/2$, with large, but finite $L_x$, $L_y$; denote
$h_{0z}^2=-\partial^2_z$ on $L^2({\mathbb{R}}, dz)$ and redefine
$h_0^2=h_\perp^2+ h_{0z}^2$. Change the model for plates by changing the
operator of $z$-motion from $h_z$, which ensures strict Dirichlet
condition, to $\tilde{h}_z$. For the moment we only assume that
$\tilde{h}_z-h_{0z}$ is a bounded operator on $L^2({\mathbb{R}}, dz)$,
commuting with the complex conjugation. Finally, set
$\tilde{h}^2=h_\perp^2+
\tilde{h}_z^2$.

With this constructions the operators $h_0^{-1}$ and $\tilde{h}^{-1}$ are
bounded, whereas the domains of $h_0$ and $\tilde{h}$ are identical (the
last statement follows from the equivalence of the norms on
${\mathcal{D}}(h_0)$: $\left(\|\psi\|^2+
\|h_0\psi\|^2\right)^{1/2}$ and $\big(\|\psi\|^2+
\|\tilde{h}\psi\|^2\big)^{1/2}$). We modify the choice of the field algebra
by replacing the original space ${\mathcal{L}}_0$ by
 $ {\mathcal{L}}={\mathcal{D}}_{\mathbb{R}}(h_0)\oplus
 L^2_{\mathbb{R}}((-L_x/2, +L_x/2)\times(-L_y/2, +L_y/2)
 \times{\mathbb{R}}, d^3x)$, where the subscript $\mathbb{R}$
denotes the real part. The defining Eqs.\ (\ref{sym},\ref{alg}) remain
intact. It is now easy to show that both the free $h_0$-evolution as well
as all new $\tilde{h}$-evolutions are correctly defined on our new
algebra by Eqs.\ (\ref{cev}--\ref{qev}) (for the $\tilde{h}$-evolutions
obvious changes of notation are to be understood: $h_0\rightarrow
\tilde{h}$, $V_0(t)\rightarrow \tilde{V}(t)$, $\alpha_{0t}\rightarrow
\tilde{\alpha}_t$). The $\tilde{h}$-evolutions are interpreted as the
evolutions of the field under the external conditions created by the
influence of the plates.

The construction of the vacuum representation of the modified algebra
remains unchanged, as outlined by Eqs.\ (\ref{j}--\ref{en}), except that
now
 ${\mathcal{K}}=L^2((-L_x/2, +L_x/2)
 \times(-L_y/2,+L_y/2) \times{\mathbb{R}}, d^3x)$, and
$j_0:{\mathcal{L}}\mapsto{\mathcal{K}}$ in (\ref{j}). By similar method
one constructs ``minimal energy state'' representations with respect to
each of the $\tilde{h}$-evolutions (``energy'' means now the field energy
together with interaction energy with fixed plates). The analog of Eq.\
(\ref{j}) defines $\tilde{\jmath}$ (well defined, as
${\mathcal{D}}(A)\subset {\mathcal{D}}(A^{1/2})$ for each positive $A$),
and the analog of Eq.\ (\ref{pos}) shows its role. The new
representations $\tilde{\pi}(\phi(V))\equiv\tilde{\Phi}(V)$ are
constructed in the same Fock space, and with the use of the same creation
and annihilation operators, but with $j_0$ replaced by $\tilde{\jmath}$
in the analog of (\ref{rep}). The $\tilde{h}$-evolution is implemented in
this representation as in (\ref{imp}) if we replace $V_0(t)$ by
$\tilde{V}(t)$, $\Phi_0$ by $\tilde{\Phi}$ and  $U_0(t)$ by
$\tilde{U}(t)=\exp i\tilde{H}t$. $\tilde{H}$ is given by (\ref{en}), with
$h_0$ replaced by $\tilde{h}$, but only up to a multiple of the unit
operator. This ambiguity becomes nontrivial if one changes the position
of the plates (and, consequently, $\tilde{h}$), and is the result of our
not having the full interacting theory at our disposal. The vector state
$\Omega_0$, however, with no ambiguity is the minimal $\tilde{H}$-energy
state in this representation.

 In the next step towards our goal one has to answer
the question, whether various constructed representations are unitarily
equivalent. If they are not, the situations to which they refer are
physically non-comparable, and no quantities referring to the change of
the distance between the plates may sensibly be determined. As the vacuum
representation $\Phi_0(V)$ defines the energy of the field itself, we
want to transform the other representations to this one. We ask
therefore, whether there does exist for each $\tilde{h}$ a unitary
operator $Q$ such that
 $Q\, \tilde{\Phi}(V)\, Q^* = \Phi_0(V)$
for all $V\in{\mathcal{L}}$. To answer the question one uses standard
methods. One can show that $j_0({\mathcal{L}})={\mathcal{D}}(h_0^{1/2})$
and $\tilde{\jmath}({\mathcal{L}})={\mathcal{D}}(\tilde{h}^{1/2})$.
Denote by $K$ the operator of complex conjugation on $\mathcal{K}$ and
define operators
 $T=2^{-1}(j_0\tilde{\jmath}^{-1}- ij_0\tilde{\jmath}^{-1}i)=
 2^{-1}(h_0^{1/2}\tilde{h}^{-1/2}+h_0^{-1/2}\tilde{h}^{1/2})$,
 $S=2^{-1}(j_0\tilde{\jmath}^{-1}+ ij_0\tilde{\jmath}^{-1}i)=
 2^{-1}(h_0^{1/2}\tilde{h}^{-1/2}-h_0^{-1/2}\tilde{h}^{1/2})K$
transforming ${\mathcal{D}}(\tilde{h}^{1/2})$ into
${\mathcal{D}}(h_0^{1/2})$. The morphism
$\tilde{\Phi}(V)\mapsto\Phi_0(V)$ may be equivalently expressed as a
Bogoliubov transformation $a(f)\mapsto b(f)\equiv a(Tf)+a^*(Sf)$, for all
$f\in{\mathcal{D}}(\tilde{h}^{1/2})$. This transformation is unitarily
implementable, $b(f)=Q\, a(f)\, Q^*$, if, and only if, the operator $S$
is a Hilbert--Schmidt operator
\cite{bra}, i.e.\ the trace of $S^*S$ is finite (hence, in particular,
$T$ and $S$ are bounded). When calculating this trace, one shows that the
summation over the degrees of freedom parallel to the plates may be
explicitly carried out, and for large dimensions of the plates (large
$L_x$ and $L_y$) one obtains
\vspace*{-.3cm}
\begin{equation}\label{eq}
 \frac{{\mathrm{Tr}}\, S^*S}{L_xL_y}\rightarrow
 \frac{\pi}{4}\, {\mathrm{Tr}}\, (\tilde{h}_z-h_{0z})^2\, ,
\end{equation}
where on the rhs the operators and the trace are regarded as operations
on $L^2({\mathbb{R}}, dz)$. Thus to satisfy our requirements we assume
that $\tilde{h}_z-h_{0z}$ is a Hilbert--Schmidt operator. We can
describe, then, all situations of interest to the Casimir effect in the
representation $\Phi_0$. In particular, the state minimizing the sum of
field energy and the energy of interaction with external conditions (the
sum given by the operator $\tilde{H}$ in the representation
$\tilde{\Phi}$), which was described by the vector $\Omega_0$ in the
representation $\tilde{\Phi}$, is given now by $\Omega=Q\Omega_0$.

Now we come to the determination of the Casimir force. In concord with
the usual treatments we assume that the states $\Omega$ (for varying
position of the plates) transform adiabatically into each other. Contrary
to implicit assumptions of most of the usual treatments, however, we
think that for the purpose of calculating actual force one should compare
the expectation value in these states of {\em the energy of the field
itself} represented by the operator $H_0$, without including the
interaction energy. We support this view by three arguments: (1) as
pointed out above, $H_0$ is the only unambiguous energy operator in the
problem, (2) in closed electrodynamics the interaction energy is absorbed
by the pure (canonical) matter energy to form the full mechanical energy
of the matter, (3) it is exactly the change in the classical analog of
$H_0$ which is used for the calculation of the force exerted on a
conductor in a classical electromagnetic field \cite{jac}. The quantity
to be calculated is, therefore,
 $(\Omega, H_0\,\Omega)=(\Omega_0, Q^*H_0 Q\,\Omega_0)$.
Using (\ref{en}) and $Q^*\, a(f)\, Q=a(T^*f)-a^*(S^*f)$ one obtains
 $(\Omega, H_0\Omega)=
 4^{-1}{\mathrm{Tr}}\,(\tilde{h}-h_0)\tilde{h}^{-1}(\tilde{h}-h_0)$.
Summing the parallel degrees of freedom one gets for large $L_x$
and~$L_y$
\begin{equation}\label{ce}
 \frac{(\Omega, H_0\Omega)}{L_xL_y}\rightarrow
 \frac{1}{24\pi}\, {\mathrm{Tr}}\,
 (\tilde{h}_z-h_{0z})(\tilde{h}_z+2h_{0z})(\tilde{h}_z-h_{0z})\, .
\end{equation}
If this is finite, the states $\Omega$ are energetically comparable, and
the Casimir force may by determined.

Finally, we specify the ``softened Dirichlet condition''. We guess that
for the appearance of some universality in the Casimir effect, as
incorporated by Casimir's original prediction, the behaviour of
$\tilde{h}_z-h_{0z}$ at the lower end of the spectrum of both $h_z$ and
$h_{0z}$ is decisive. We put, therefore
 $\tilde{h}_z-h_{0z}=f(h_z)-f(h_{0z})$,
where $f$ is a real smooth function on ${\mathbb{R}}_+$, with $f(u)=u$
for small $u$, $0\leq f(u)\leq u$ for all $u$,  and vanishing at least as
$u^{-2}$ for $u\rightarrow\infty$. This ensures finiteness of both
(\ref{eq}) and (\ref{ce}). The resulting $\tilde{h}$ will not guarantee
the relativistic causality of the evolution, but for the quasi-static
idealization this is not a serious objection. Denote the Casimir energy
(\ref{ce}) for this model by ${\mathcal{E}}(a)$, and introduce the
abbreviation
 $\chi(\kappa, p)=\kappa^2(\kappa^2-p^2)^{-2}
 (f(\kappa)-f(p))^2(3p-f(p)+f(\kappa))$.
Then
\vspace*{-.2cm}
\begin{eqnarray}\label{ea}
 {\mathcal{E}}(a)=&&\frac{1}{6\pi^3}\sum_{k=1}^\infty\frac{\pi}{a}
 \int_0^\infty dp\,\chi\Big(\frac{k\pi}{a}, p\Big)
 \left(1+(-1)^{k+1}\cos ap\right) \nonumber\\
 +&&\frac{1}{6\pi^3}\int_0^\infty d\kappa\int_0^\infty dp\,
 \chi\left( \kappa, p\right)\, .
\end{eqnarray}
A rather lengthy analysis of this expression shows that
\begin{equation}\label{CE}
 {\mathcal{E}}(a)={\mathcal{E}}(\infty)-\frac{\pi^2}{1440}\, a^{-3} +
 o(a^{-3})\, ,
\end{equation}
\begin{equation}
 \label{CF}
 -\frac{d{\mathcal{E}}}{da}(a)=-\frac{\pi^2}{480}\, a^{-4}
 + o(a^{-4})\, .
\end{equation}
One recognizes in these expressions the familiar Casimir terms -- the
leading asymptotic  term in the force and the second leading term in the
energy (which are one half of the corresponding terms for the
electromagnetic Casimir quantities). They are here determined completely
by the behaviour of the function $f$ in the neighborhood of zero (in
fact, the property $f'(0)=1$ is all one needs). However, their meaning
here is different, the force obeys this simple law only for sufficiently
large separations of the plates. For every finite separation other
($f$-dependent) terms will dominate for $f$ approaching identity. It
becomes evident from (\ref{ea}) that reaching this limit is both
physically and mathematically meaningless -- the energy becomes infinite
and the force indeterminate. Observe, also, that here the energy
${\mathcal{E}}(a)$ is always positive, as it should be. The physical
interpretation of the $f$-dependent limit ${\mathcal{E}}(\infty)$ is the
following: it equals twice the work which the external forces have to
perform to create the configuration of the field surrounding a single
plate (in this limit the plates and the configurations around them may be
regarded as independent). One checks the consistency of this
interpretation by repeating the calculation for the configuration with
only one plate present in the whole space, and finding that the resulting
energy is indeed one half of ${\mathcal{E}}(\infty)$.

The calculation of the effect as performed here used a class of models
determined by the function $f$. However, as mentioned above, the leading
term in the force comes from the spectral area where $\tilde{h}_z=h_z$,
so it is probably more universal. At the same time Eq.\ (\ref{ce}) gives
the method for the construction of other models, and corrections to the
leading terms.

Lessons to be drawn for experimental verification of the Casimir effect
are as follows: first, the universality is to be searched for at large
separation of the plates, and second, for smaller separations
model-dependent aspects take over.

I am grateful to D. Buchholz for helpful discussions.

\end{document}